\documentclass[11pt]{article}
\usepackage{amsfonts}
\usepackage{amssymb}
\usepackage{amsmath,amssymb}
\usepackage{bbm}
\usepackage{tikz}

\def\slasha#1{\setbox0=\hbox{$#1$}#1\hskip-\wd0\hbox to\wd0{\hss\sl/\/\hss}}

\def\periodb#1{\setbox0=\hbox{$#1$}#1\hskip-\wd0\hbox to\wd0{-}}

\def\sfrac#1#2{{\textstyle\frac{#1}{#2}}}
\newcommand{\unit}{\mathbbm{1}}   

\newcommand{\CJ}{\mathcal{J}} 
\newcommand{\CT}{\mathcal{T}} 
    
\newcommand{\CF}{\mathcal{F}}   
\newcommand{\CG}{\mathcal{G}}  
\newcommand{\CH}{\mathcal{H}}

\newcommand{\CV}{\mathcal{V}} 
   
\newcommand{\R}{\mathbb{R}}     
\newcommand{\C}{\mathbb{C}}     

\newcommand{\im}{\mathrm{i}} 
 
\newcommand{\dd}{\mathrm{d}}     
\newcommand{\dpar}{\partial}     
\newcommand{\diag}{{\mathrm{diag}}}

\newcommand{\+}{\dagger}

\newcommand{\Ad}{\mathrm{Ad}} 
\newcommand{\sU}{\mathrm{U}}     
\newcommand{\sSU}{\mathrm{SU}}     
     
\newcommand{\sSO}{\mathrm{SO}}     
\newcommand{\sGL}{\mathrm{GL}}  
\newcommand{\sSp}{\mathrm{Sp}}

\newcommand{\ru}{\mathrm{u}}    
\newcommand{\rS}{\mathrm{S}} 

\newcommand{\al}{{{\alpha}}}

\newcommand{\vk}{{{\varkappa}}}

\newcommand{\xh}{\hat{x}}

\newcommand{\bb}{{\bar{b}}}
\newcommand{\yb}{{\bar{y}}}
\newcommand{\zb}{{\bar{z}}}
\newcommand{\mb}{{\bar{\mu}}}
\newcommand{\nb}{{\bar{\nu}}}

\newcommand{\Av}{{A_{\sf{vac}}}}
\newcommand{\Fv}{{F_{\sf{vac}}}}
\newcommand{\Ev}{{E^{\sf{v}}}}

\newcommand{\qv}{{q_{\sf{v}}}}

\newcommand{\und}{\quad\mbox{and}\quad}
\newcommand{\with}{\quad\mbox{with}\quad}
\newcommand{\for}{\quad\mbox{for}\quad}

\makeatletter

\@addtoreset{equation}{section}
\makeatother

\begin{document}
\begin{titlepage}
\setcounter{page}{0}
.
\vskip 15mm
\begin{center}
{\LARGE \bf Klein-Gordon oscillators and Bergman spaces}\\
\vskip 2cm
{\Large Alexander D. Popov}
\vskip 1cm
{\em Institut f\"{u}r Theoretische Physik,
Leibniz Universit\"{a}t Hannover\\
Appelstra{\ss}e 2, 30167 Hannover, Germany}\\
{Email: alexander.popov@itp.uni-hannover.de}
\vskip 1.1cm
\end{center}
\begin{center}
{\bf Abstract}
\end{center}

\noindent 
We consider classical and quantum dynamics of relativistic oscillator in Minkowski space $\mathbb{R}^{3,1}$. It is shown that for a non-zero frequency parameter $\omega$ the covariant phase space of the classical Klein-Gordon oscillator is a homogeneous K\"ahler-Einstein manifold $Z_6=\mathrm{Ad}S_7/\mathrm{U}(1)=\mathrm{U}(3,1)/\mathrm{U}(3)\times \mathrm{U}(1)$. In the limit $\omega\to 0$, this manifold is deformed into the covariant phase space $T^*H^3$ of a free relativistic particle, where $H^3=H^3_+\cup H_-^3$ is a two-sheeted hyperboloid in momentum space. Quantization of this model with $\omega\ne 0$ leads to the Klein-Gordon oscillator equation which we consider in the Segal-Bargmann representation. It is shown that the general solution of this model is given by functions from the weighted Bergman space of square-integrable holomorphic (for particles) and antiholomorphic (for antiparticles) functions on the K\"ahler-Einstein manifold $Z_6$. This relativistic model is Lorentz covariant, unitary and does not contain non-physical states.

\end{titlepage}

\section{Introduction}

The aim of this paper is to consider the possibility of eliminating the problems of relativistic quantum mechanics (negative energies, negative norm states, etc.) using the example of relativistic harmonic oscillator. It is defined as a particle of mass $m$ in Minkowski space $\R^{3,1}$ with coordinates $x^\mu$ and momenta $p_\mu$, $\mu =0,...,3$, in a field of external forces specified by the function $V(x)=m^2\omega^2\eta_{\mu\nu}x^\mu x^\nu$. Here $\omega$ is the angular frequency and $(\eta_{\mu\nu}^{})=\mbox{diag}(-1,1,1,1)$ is the Minkowski metric. Quantization of this model leads to the Klein-Gordon oscillator equation,
\begin{equation}\label{1.1}
\Bigl (\eta^{\mu\nu}\frac{\dpar}{\dpar x^\mu}\frac{\dpar}{\dpar x^\nu} -m^2 -
m^2\omega^2\eta_{\mu\nu}x^\mu x^\nu\Bigr )\Psi =0\ ,
\end{equation}
considered in many papers (see e.g.~\cite{Dirac}-\cite{Bedic} and references therein). This equation can be viewed as a deformation of the free Klein-Gordon equation corresponding to the limit $\omega\to 0$. The solution space of equation \eqref{1.1} was considered e.g. in \cite{Bars, Bedic}. It has been shown that this equation admits ground state of the form
\begin{equation}\label{1.2}
\psi_0^\pm =\exp \Bigl(\pm\frac{m\omega}{2}\eta_{\mu\nu}x^\mu x^\nu \Bigr)\ ,
\end{equation}
\begin{equation}\label{1.3}
\psi_0 =\exp \Bigl(-\frac{m\omega}{2}\delta_{\mu\nu}x^\mu x^\nu \Bigr)\ .
\end{equation}
It was argued that in cases \eqref{1.2} Lorentz covariance holds but there are an infinite number of non-physical states, whereas in case \eqref{1.3} there are no non-physical states but Lorentz covariance is violated.

The last statement seems to be erroneous since Dirac showed \cite{Dirac} as early as 1945 that equation \eqref{1.1} can be rewritten as
\begin{equation}\label{1.4}
\Bigl(\xi^\mu\frac{\dpar}{\dpar\xi^\mu} + N +2\Bigr)\Psi =0\ ,
\end{equation}
where 
\begin{equation}\label{1.5}
\xi^\mu :=\frac{1}{\sqrt{2}}\Bigl( x^\mu - \frac{1}{m\omega}\eta^{\mu\nu}\frac{\dpar}{\dpar x^\nu}\Bigr)\ ,\quad \frac{\dpar}{\dpar\xi^\mu}:=\frac{1}{\sqrt{2}}\Bigl(\frac{\dpar}{\dpar x^\mu}+m\omega\eta_{\mu\nu}x^\nu\Bigr)
\end{equation}
and $N:=m/2\omega =1,2,...$ is an integer given by the choice of $\omega$. Formal solutions of equation \eqref{1.4} have the form
\begin{equation}\label{1.6}
\Psi (n_0,n_1,n_2,n_3)=\frac{1}{(\xi^0)^{n_0}}(\xi^1)^{n_1}(\xi^2)^{n_2}(\xi^3)^{n_3}\with n_0{-}n_1{-}n_2{-}n_3=N{+}2
\end{equation}
and the space of such solutions forms an infinite-dimensional unitary representation of the Lorentz group SO(3,1). Dirac also showed that solutions \eqref{1.6} are in one-to-one correspondence with solutions
\begin{equation}\label{1.7}
\Bigl(\frac{\dpar}{\dpar\xi^0}\Bigr)^{n_0-1}     (\xi^1)^{n_1}(\xi^2)^{n_2}(\xi^3)^{n_3}\psi_0
\end{equation}
in the model with the ground state \eqref{1.3}, as are their inner products (unitary equivalence), and hence this model is also Lorentz covariant \cite{Dirac}.

Despite the many citations of Dirac's paper \cite{Dirac}, it appears that his ideas were not accepted and developed. This may be due to the fact that in the eigenstates \eqref{1.6} the operator $\xi^0$ appears in negative powers, which makes it difficult to interpret these eigenstates. In this paper we will show that this problem disappears if we move from the position representation \eqref{1.1}-\eqref{1.7} to the complex Bargmann-Fock-Segal representation of the canonical commutation relations for the relativistic oscillator. We will show that the operator form of $\xi^\mu$ in \eqref{1.4}-\eqref{1.7} and other problems in the coordinate representation are related to the fact that the covariant phase space of the Klein-Gordon oscillator is a homogeneous space
\begin{equation}\label{1.8}
Z_6=\Ad S_7/\sU(1)\cong\sSU(3,1)/\rS(\sU(3)\times \sU(1))
\end{equation}
parametrized by complex coordinates
\begin{equation}\label{1.9}
y^a=\frac{z^a}{z^0}\with z^\mu :=\frac{1}{\sqrt 2}\Bigl (x^\mu - \frac{\im}{m\omega}\eta^{\mu\nu}p_\nu\Bigr )\ ,\quad a=1,2,3,
\end{equation}
which is incompatible with the choice of functions depending only on coordinates or momenta.

When using complex coordinates $z^\mu$ on the phase space $T^*\R^{3,1}\cong\C^{3,1}$, the Klein-Gordon oscillator equation in the holomorphic representation is reduced to the equation
\begin{equation}\label{1.10}
\Bigl(z^\mu\frac{\dpar}{\dpar z^\mu}+N+2\Bigr)\Psi =0\  ,
\end{equation}
where, in contrast to \eqref{1.4}, $z^\mu$ are ordinary coordinates. One of the goals of this paper is to give a clear geometric meaning to Dirac's idea of an infinite-dimensional unitary representation of the Lorentz group on solutions of the Klein-Gordon oscillator equation. 
Namely, we will show that solutions of equation \eqref{1.10} are parametrized by the weighted Bergman space
\begin{equation}\label{1.11}
L_h^2(Z_6, \mu_{N+2})\with \mu^{}_{N+2}=(1-\delta_{a\bb}\,y^a\yb^\bb)^{N+2}\ ,
\end{equation}
which is the space of square-integrable holomorphic functions on $Z_6$ with a weight function $\mu^{}_{N+2}$ in the inner product. The group SO(3,1) is embedded in the group $\sSU(3,1)$ acting on the K\"ahler-Einstein coset space $Z_6$ and on the Hilbert space $L_h^2(Z_6, \mu^{}_{N+2})$ realizing a unitary representation of SO(3,1). In this way we describe an exactly solvable unitary Lorentz covariant model of quantum Klein-Gordon oscillator. Using this model as a guiding example, we see that the solution space of the classical model is the space of initial data \eqref{1.8}, and the solution space of the quantum model is the polarized space \eqref{1.11} of functions on the space of initial data $Z_6$ of the classical model. If we apply this observation to free particles, then the solution space of the Klein-Gordon equation should be parametrized by the space of functions on the hyperboloid $H^3$ in the $p$-space satisfying the real analogue of equation \eqref{1.10} of type $(p_\mu\dpar/\dpar p_\mu +2)\psi (p)=0$. This will lead to a real analogue of the Bergman space \eqref{1.11}, which requires further study.

\section{Free relativistic particles}

\noindent {\bf Symplectic structure}.  Let us consider the phase space $T^*\R^{3,1}\cong\R^{6,2}$ of relativistic spinless particles with coordinates $x^\mu\in\R^{3,1}$ and momenta $p_\mu\in\R^{3,1}$, $\mu , \nu =0,...,3$. The canonical symplectic structure on $T^*\R^{3,1}$ is
\begin{equation}\label{2.1}
\omega_{\R^{6,2}}^{}=\dd p_\mu\wedge\dd x^\mu =\omega_{\mu\,\nu+4}\dd x^\mu\wedge\dd x^{\nu +4}=\frac{1}{w^2}\,\eta_{\mu\nu}\dd x^\mu\wedge\dd x^{\nu +4}\ ,
\end{equation}
where $w\in\R^+$ is a length parameter, so that $[w^2p_\mu]=[\mbox{length}]=[x^\mu]$ for $\hbar =c=1$, 
\begin{equation}\label{2.2}
x^{\mu +4}:=-w^2p^\mu = -w^2\eta^{\mu \nu}p_\nu\und \omega_{\mu\,\nu+4}=\frac{1}{w^2}\,\eta_{\mu\nu}=-\omega_{\nu +4\,\mu}\ .
\end{equation}
The bivector field inverse to the two-form \eqref{2.1} has components
\begin{equation}\label{2.3}
\omega^{\mu\,\nu+4}=-w^2\eta^{\mu\nu}=-\omega^{\nu+4\, \mu}
\end{equation}
so that
\begin{equation}\label{2.4}
\omega^{\mu\,\sigma+4}\omega_{\sigma+4\,\nu}=\delta_\nu^\mu\und
\omega^{\mu+4\,\sigma}\omega_{\sigma\,\nu+4}=\delta_\nu^\mu\ .
\end{equation}
The two-form $\omega_{\R^{6,2}}^{}$ is non-degenerate.

Classical relativistic particle is a point in $T^*\R^{3,1}$ moving along a trajectory defined by a Hamiltonian vector field
\begin{equation}\label{2.5}
V_H^{}=\omega^{\mu+4\,\nu}\dpar_{\mu+4}H\dpar_\nu + \omega^{\mu\,\nu+4}\dpar_\mu H\dpar_{\nu +4}\with
\dpar_\mu :=\frac{\dpar}{\dpar x^\mu},\ \dpar_{\mu +4}:=\frac{\dpar}{\dpar x^{\mu +4}}\ ,
\end{equation}
where $H=H(x,p)$ is a function (Hamiltonian) on the phase space $T^*\R^{3,1}$ that specifies the dynamics. 
This conserved quantity $H$ is not an energy function of the particle.
The particle moves in phase space along a trajectory that is determined from Hamiltonian flow equations
\begin{equation}\label{2.6}
\dot x^\mu = V_H^{} x^\mu\und \dot x^{\mu +4}= V_H^{} x^{\mu +4}\ ,
\end{equation}
where $\dot x = \dd x/\dd\tau$ and $\tau$ is the proper time.

\noindent {\bf Metric tensor}.  Recall that symplectic structures on the space $\R^8$ are parametrized by the homogeneous space $\sGL(8,\R)/\sSp(8,\R)$. Choice \eqref{2.1} fixes the subgroup $\sSp(8,\R)$ in the Lie group $\sGL(8,\R)$ of all linear transformations of the space $\R^8$ \cite{Kob}. Similarly, metrics of signature (6,2) on the space $\R^8$ are parametrized by the homogeneous space $\sGL(8,\R)/\sSO(6,2)$ and we choose the canonical diagonal metric
\begin{equation}\label{2.7}
g^{}_{\R^{6,2}}=\eta_{\mu\nu}\dd x^\mu\dd x^\nu + w^4\eta^{\mu\nu}\dd p_\mu\dd p_\nu =\eta_{\mu\nu}(\dd x^\mu\dd x^\nu + \dd x^{\mu +4}\dd x^{\nu +4})=:-\dd\tau^2
\end{equation}
with the inverse metric
\begin{equation}\label{2.8}
g^{-1}_{\R^{6,2}}=\eta^{\mu\nu}\frac{\dpar}{\dpar x^\mu}\otimes \frac{\dpar}{\dpar x^\nu} + \frac{1}{w^4}\,\eta_{\mu\nu}
\frac{\dpar}{p_\mu}\otimes \frac{\dpar}{\dpar p_\nu}
=\eta^{\mu\nu}(\dpar_\mu\otimes\dpar_\nu + \dpar_{\mu +4}\otimes\dpar_{\nu +4})=:-\dpar_\tau\otimes\dpar_\tau\ .
\end{equation}
Thus we have components 
\begin{equation}\label{2.9}
g_{\mu\nu}=\eta_{\mu\nu}=g_{\mu+4\,\nu+4}\und g^{\mu\nu}=\eta^{\mu\nu}=g^{\mu+4\,\nu+4}
\end{equation}
when using the basis $\{\dd x^\mu , \dd x^{\mu +4}\}$ for one-forms on $\R^8$ and the basis $\{\dpar_\mu , \dpar_{\mu +4}\}$ for vector fields. In \eqref{2.7} we introduced an affine parameter $\tau$ on the trajectory of particle in the phase space. It depends not only on its relative velocity but also on its acceleration. Note that the evolution parameter $\tau$ is a scalar for transformations of the pseudo-unitary group
\begin{equation}\label{2.10}
\sU(3,1)=\sSp(8, \R)\cap\sSO(6,2)
\end{equation}
preserving both the metric \eqref{2.7} and the symplectic 2-form \eqref{2.1}. Lorentz group SO(3,1) is an obvious subgroup in the group \eqref{2.10}. Note that the metric \eqref{2.7} and its inverse \eqref{2.8} admit a limit $w^2\to\infty$ in which momenta $p_\mu$ become constant (free particles) so that $\dd p_\mu =0$. In this case, $\tau$ becomes the usual proper time for a free particle in Minkowski space. This is a degenerate case.

\noindent {\bf Free particles: equations of motion}.  Free massive particles are specified by the Hamiltonian
\begin{equation}\label{2.11}
H=-\frac{1}{2m}\,\eta^{\mu\nu}p_\mu p_\nu =-\frac{1}{2m}\,\eta_{\mu\nu}p^\mu p^\nu \equiv 
-\frac{1}{2mw^4}\,g_{\mu +4\,\nu +4}x^{\mu +4}x^{\nu +4}\ ,
\end{equation}
for which the vector field \eqref{2.5} has the form
\begin{equation}\label{2.12}
V_H^{} = -\frac{1}{mw^2}\,x^{\mu +4}\dpar_\mu =\frac{p^\mu}{m}\,\dpar_\mu \for p^\mu =\eta^{\mu\nu}p_\nu =: mv^\mu\ .
\end{equation}
For the vector field \eqref{2.12} equations \eqref{2.6} and their solutions have the form
\begin{equation}\label{2.13}
\dot x^\mu = v^\mu\ ,\quad \dot v^\mu =0\ \Rightarrow\ x^\mu (\tau )=x^\mu + v^\mu\tau\ .
\end{equation}
Substituting solution \eqref{2.13} into \eqref{2.7} we get
\begin{equation}\label{2.14}
\eta_{\mu\nu}\,\frac{\dd x^\mu}{\dd\tau}\frac{\dd x^\nu}{\dd\tau}\dd\tau^2 = \frac{1}{m^2}\, 
 \eta_{\mu\nu}\,p^\mu p^\nu\dd\tau^2=-\dd\tau^2\quad\Rightarrow\quad\eta_{\mu\nu}\,p^\mu p^\nu =-m^2\ ,
\end{equation}
i.e. a free massive scalar particle moves along a timelike path\footnote{To describe massless particles, we should replace $\tau$ in \eqref{2.13} with an affine parameter $\sigma$ not related with the metric \eqref{2.7}
and again obtain \eqref{2.14} with $\dpar_\sigma$ instead of $\dpar_\tau$. Then for $m^2\to 0$ we get $\eta_{\mu\nu}p^\mu p^\nu =0$ and lightlike worldline.} in Minkowski space with constant velocity defined by the standard energy-momentum relation \eqref{2.14}.

\noindent {\bf Free particles: phase space}. To see the geometry behind \eqref{2.11}-\eqref{2.14}, we consider a map $\mu_H^{}$ (momentum map \cite{MW}) from $T^*\R^{3,1}$ to $\R$,
\begin{equation}\label{2.15}
\mu_H^{}: T^*\R^{3,1}\to \R\with \mu_H^{}(x,p)=2H=-\frac{1}{m}\,\eta_{\mu\nu}\,p^\mu p^\nu \in\R\ .
\end{equation}
The constant value $m>0$  of this function defines a hypersurface (level surface) in $T^*\R^{3,1}$,
\begin{equation}\label{2.16}
\mu_H^{-1}(m) = H^3\times\R^{3,1}=\bigl\{x,p\in T^*\R^{3,1}\mid \eta_{\mu\nu}\,p^\mu p^\nu =-m^2\bigr\}\ ,
\end{equation}
where $H^3 = H^3_+\cup H^3_-$ is the two-sheeted hyperboloid in the momentum space,
\begin{equation}\label{2.17}
H^3_+:\ \  p^0=\sqrt{\delta_{ab}p^ap^b + m^2}\und H^3_-:\ \  p^0=-\sqrt{\delta_{ab}p^ap^b + m^2}\ ,
\end{equation}
and $\R^{3,1}$ in \eqref{2.16} is the space of coordinates $x^\mu$.

On the manfold \eqref{2.16} the action of the one-parameter group
\begin{equation}\label{2.18}
\sGL(1, \R)=\R^*=\bigl\{g=\exp (\tau V_H)=\exp (\tau v^\mu\dpar_\mu )\bigr\}
\end{equation}
generated by the vector field \eqref{2.12} is given. Solutions \eqref{2.13} describe the orbits of this group in the manifold \eqref{2.16},
\begin{equation}\label{2.19}
gx^\mu= x^\mu (\tau), \ gp^\mu = p^\mu\quad \Rightarrow\quad \dot x^\mu (\tau ) =\dot gx^\mu =\frac{p^\mu}{m}\und\dot p^\mu =0\ .
\end{equation}
Quotienting by the action of the dynamics groups \eqref{2.18} is a covariant phase space (space of initial data),
\begin{equation}\label{2.20}
X^6:=\mu_H^{-1}(m)\slash\R^* = H^3\times\R^{3,1}\slash\R^*\cong T^*H^3_+\cup T^*H^3_-\ ,
\end{equation}
that parametrizes the orbits of this group. In \eqref{2.20}, the six-dimensional manifolds $T^*H^3_\pm$ are cotangent bundles over hyperboloids \eqref{2.17}, corresponding to particles and antiparticles. Manifold \eqref{2.20} is the {\it space of initial data} for equations \eqref{2.13} and \eqref{2.19}. Note that the energy of particles $(\qv =1)$ and antiparticles $(\qv =-1)$ is equal to $E=\qv p^0$ and is always positive. Here $\qv=\pm 1$ define an orientation on $\tau$-axis.

\noindent {\bf Nonrelativistic limit}.  The energy-momentum relations \eqref{2.16} and \eqref{2.17} can be used to pass to the nonrelativistic limit. Namely, restoring the speed of light $c$, we obtain
\begin{equation}\label{2.21}
E=\qv p^0c=\sqrt{\delta^{ab}p_ap_bc^2 + m^2c^4}=mc^2\sqrt{1+\frac{p^2}{m^2c^2}}\cong mc^2+\frac{p^2}{2m}\quad \mbox{for}\quad c\to\infty\ .
\end{equation}
In this case, the phase space $T^*\R^{3,1}$ is reduced to the Euclidean space $T^*\R^{3}=\R^6$ with an obvious symplectic structure and metric \cite{Popov1}
\begin{equation}\label{2.22}
g_{\R^6}^{}=\delta_{ab}\,\dd x^a\dd x^b + w^4\delta^{ab}\dd p_a\dd p_b\ .
\end{equation}
Repeating the consideration of  \eqref{2.15}-\eqref{2.20} with $H=p^2/2m$, we obtain the phase space
\begin{equation}\label{2.23}
\mu_H^{-1}(E'=E-mc^2=const)\slash\R^* \cong T^*S^2\ ,
\end{equation}
where the 2-sphere $S^2$ is defined by the equations $p^2=2mE'$. Note also that if a particle has an electric charge, then introducing the interaction of this particle with the electric potential in \eqref{2.21} by replacing $p_0\mapsto p_0+A_0$ with $A_0\sim 1/r$, $r^2=\delta_{ab}x^ax^b$, we obtain a deformation of the phase space \eqref{2.23} into the space (see e.g. \cite{Hurt})
\begin{equation}\label{2.24}
S^2\times S^2\subset T^*\R^{3}
\end{equation}
for which the energy is constant. After quantization, we will obtain a description of the hydrogen atom and its energy levels.

\section{Klein-Gordon oscillator}

\noindent {\bf Hamiltonian function}.  Let us consider a particle of mass $m$ under the influence of an attractive force $F^\mu = -\omega^2 x^\mu$ in Minkowski space $\R^{3,1}$. This particle will perform sinusoidal oscillations around the equilibrium point $x_0^\mu$ not only in space $\R^3$ but also along the coordinate time axis $x^0\in\R$ with constant amplitude and constant frequency $\omega =2\pi/T$, where $T$ is the time for a single oscillation. If we consider large $T$, for example comparable to the age of the universe, then relativistic oscillators will be indistinguishable from free particles. However, even a very small $\omega = 2\pi/T$ changes the geometry of the covariant phase space of particle and instead of a disconnected manifold \eqref{2.20} having an infinite volume, we obtain a simply connected manifold \eqref{1.8} having a finite volume without limiting the absolute values of the coordinates and momenta. At the quantum level, the difference becomes even greater. In particular, all reasonable integrals over the covariant phase space \eqref{1.8} are finite. We will call a massive particle in a force field $F^\mu =-\omega^2 x^\mu$ a Klein-Gordon oscillator.

The Klein-Gordon oscillator is defined by the Hamiltonian function
\begin{equation}\label{3.1}
H=-\frac{1}{2mw^4} \, \eta_{\mu\nu} (x^\mu x^\nu + x^{\mu +4}x^{\nu +4})=-\frac{1}{2m}\,(\eta^{\mu\nu}p_\mu p_\nu + m^2\omega^2\eta_{\mu\nu}x^\mu x^\nu )\ ,
\end{equation}
where
\begin{equation}\label{3.2}
\omega :=\frac{1}{mw^2}\quad\Rightarrow\quad w^2=\frac{1}{m\omega}\ .
\end{equation}
The expressions in parentheses are given using the metric \eqref{2.7} and are therefore invariant with respect to group SO(6,2) and its Lorentz subgroup SO(3,1). Note that taking the limit $w^2\to\infty$ ($\omega\to 0$) we can return to the Hamiltonian \eqref{2.11} of a free particle. This possibility exists at all stages of further consideration.

\noindent {\bf Equations of motion}.  Vector field \eqref{2.5} for Hamiltonian \eqref{3.1} has the form
\begin{equation}\label{3.3}
V_H^{}=\frac{p^\mu}{m}\,\frac{\dpar}{\dpar x^\mu} - m\omega^2x^\mu\frac{\dpar}{\dpar p^\mu} =\omega (x^\mu\dpar_{\mu +4}-x^{\mu +4}\dpar_\mu )\ .
\end{equation}
This vector field is a generator of group SO(2) of rotations in planes $(x^\mu , x^{\mu +4})$. When $w^2\to\infty$ ($\omega\to 0$) \eqref{3.3} is reduced to the vector field \eqref{2.12} of a free particle. Equations of motion \eqref{2.6} for Hamiltonian \eqref{3.1} have the form
\begin{equation}\label{3.4}
\begin{split}
\dot x^\mu &= V_H^{} x^\mu =-\omega x^{\mu +4} =\frac{p^\mu}{m}\ ,\quad 
\dot x^{\mu +4} = V_H^{} x^{\mu +4}=\omega x^\mu\\
&\Rightarrow\ \ddot x^\mu +\omega^2x^\mu =0\ ,\quad \ddot p^\mu +\omega^2 p^\mu =0
\end{split}
\end{equation}
and their solutions are
\begin{equation}\label{3.5}
x^\mu (\tau ) = x^\mu\cos\omega\tau + v^\mu\frac{\sin\omega\tau}{\omega}\ ,\quad
p^\mu (\tau ) = p^\mu\cos\omega\tau - m\omega x^\mu\sin\omega\tau\ ,
\end{equation}
where $x^\mu =x^\mu(0)$ and $p^\mu=p^\mu (0)=m v^\mu=m v^\mu (0)$ are the initial data. 
The example of the relativistic oscillator \eqref{3.5} shows that the evolution parameter $\tau$ can coincide with the coordinate time $x^0$ only for free particles.
Note that the microscopic causality principle should be applied to the proper time $\tau$, and not to the coordinate time $x^0$.

\noindent {\bf Level set.} Substituting solution \eqref{3.5} into \eqref{2.7} with $\dd x^\mu =\dot x^\mu\dd\tau$ and $\dd x^{\mu +4}=\dot x^{\mu +4}\dd\tau$, we obtain the constraint equation
\begin{equation}\label{3.6}
\eta^{\mu\nu}p_\mu p_\nu + \frac{1}{w^4}\eta_{\mu\nu}x^\mu x^\nu =-m^2
\end{equation}
which defines the level set $\mu_H^{-1}(m)$ for the momentum map
\begin{equation}\label{3.7}
\mu_H^{} : T^*\R^{3,1}\to\R\with \mu_H^{}=2H=-\frac{1}{m}\,\eta_{\mu\nu}(p^\mu p^\nu + \frac{1}{w^4}\, x^\mu x^\nu )\in\R^+\ .
\end{equation}
Thus, the oscillating particle of mass $m$ for any value of proper time $\tau$ is located on the hypersurface \eqref{3.6} in space $T^*\R^{3, 1}$. This is a level surface of the Hamiltonian \eqref{3.1}. In components we have 
\begin{equation}\label{3.8}
E^2:= p_0^2 + \left(\frac{x^0}{w^2}\right)^2 = \delta^{ab}p_a p_b +\frac{1}{w^4}\,\delta_{ab}x^ax^b + m^2\ ,
\end{equation}
where $a,b=1,2,3$. From  \eqref{3.8} it follows that the energy $E$ coincides with the radius of the circles $S^1$ in the $(x^0, p_0)$-plane and therefore it cannot be negative. It is also obvious that in the limit $w^2\to\infty$ we will get two points, $p^0=E$ and $p^0=-E$ with $E>0$. 
This can also be written as $p^0=\qv E$, where  $\qv =\pm 1$ correspond to two directions on the $\tau$-axis (orientation).

Note that the level surface \eqref{3.6} does not depend on $\tau$ and defines the anti-de Sitter space Ad$S_7$. Independence of $\tau$ means that group U(1) with generator $V_H$ from \eqref{3.3} maps this manifold into itself, i.e. 
\begin{equation}\label{3.9}
\sU(1) \ni g=e^{\tau V_H}: \quad (x^\mu , p^\mu ) \mapsto (x^\mu (\tau ), p^\mu (\tau ))\in \Ad S_7\ .
\end{equation}
Hence, Ad$S_7$ is the total space of the U(1)-bundle
\begin{equation}\label{3.10}
\Ad S_7\ \stackrel{S^1}{\longrightarrow}\ \Ad S_7/\sU(1)\cong \sU(3,1)/\sU(3)\times \sU(1)
\end{equation}
and the space $\Ad S_7/\sU(1)$ of orbits of group U(1) is parametrized by the space of initial data at $\tau =0$, i.e. is a covariant phase space of Klein-Gordon oscillator.

\noindent {\bf Nonrelativistic limit.} The Ad$S_7$ level surface equation \eqref{3.8} can be used to pass to the nonrelativistic limit in the same way as was done in \eqref{2.21}-\eqref{2.23} for free particles. To do this, first in the metric \eqref{2.7} you should rescale $p_0\mapsto \gamma p_0$ to get $w^4_0\dd p_0^2:=\gamma^2 w^4 \dd p^2_0$, which will lead to replacing $x^0/w^2$ with
$x^0/w_0^2$ in \eqref{3.8}. After this, we can take the limit $\gamma^2\to\infty$ and obtain the constraint equation
\begin{equation}\label{3.11}
E^2 = p^2_0 =\delta^{ab}p_a p_b + \frac{1}{w^4}\,\delta_{ab}x^ax^b + m^2
\end{equation}
defining the level surface $H^6\times\R$ in $T^*\R^{3,1}$, where $H^6 =H_+^6\cup H_-^6$ is the two-sheeted hyperboloid and $\R$ is parametrized by $x^0$. Restoring the speed of light in \eqref{3.11} we obtain
\begin{equation}\label{3.12}
E=\qv p^0c=\sqrt{(\delta^{ab}p_ap_b+ \frac{1}{w^4}\,\delta_{ab}x^ax^b)c^2+m^2c^4}\cong mc^2+\frac{p^2}{2m}+\frac{m\omega^2}{2}\, x^2\quad \mbox{for}\quad c\to\infty\ .
\end{equation}
Thus, we obtain the standard nonrelativistic oscillator with $E'=E-mc^2$. Note that $p^0=\qv E/c$ with $E>0$ for $\qv =1$ (particles) and $\qv =-1$ (antiparticles).

\noindent {\bf Complex structures.} Consider space $\R^8$ on which the general linear group $\sGL(8,\R$) acts via a change of basis. The introduction of a symplectic 2-form \eqref{2.1} on $\R^8$ reduces the group $\sGL(8,\R$) to a subgroup $\sSp(8,\R$) of transformations preserving this 2-form. By introducing metric \eqref{2.7} of signature (6,2) on $\R^8$, we fix in $\sGL(8,\R$) a subgroup SO(6,2) preserving this metric. If we want to preserve both the metric and the symplectic structure at the same time, then we should reduce 
$\sGL(8,\R$) to the subgroup U(3,1) from \eqref{2.10}. The introduction of an almost complex structure $\CJ$ on $\R^8$ reduces the group $\sGL(8,\R$) to a subgroup $\sGL(4,\C$) preserving $\CJ$ \cite{Kob}. Any such complex structure on $\R^8$ is admissible and is parametrized by the homogeneous space $\sGL(8, \R)/\sGL(4,\C$). If we want this complex structure is consistent with both the metric and the symplectic form, then it is necessary that it be preserved by the subgroup U(3,1) of the group $\sGL(4,\C$). This holds if $\CJ$ is given by the formulae
\begin{equation}\label{3.13}
\CJ_N^M=w^2g^{MK}\omega_{NK}\for M,N =(\mu , \mu +4)\ .
\end{equation}
For the tensors $\omega^{}_{\R^{6,2}}$ and $g^{}_{\R^{6,2}}$ we introduced, we obtain
\begin{equation}\label{3.14}
\CJ_\mu^{\nu +4}=\delta_\mu^\nu\und \CJ^\mu_{\nu +4}=-\delta^\mu_\nu\ \Rightarrow\ \CJ_K^M\CJ_N^K=-\delta_N^M
\end{equation}
in the basis $\dpar/\dpar x^M$ and $\dd x^M$. The complex structure \eqref{3.14} is invariant under transformations of group U(3,1) by construction. 

The complex structure \eqref{3.14} defines on $\R^8$ complex coordinates
\begin{equation}\label{3.15}
z^\mu =\frac{1}{\sqrt 2}(x^\mu + \im x^{\mu +4})\und \zb^{\bar \mu}=\frac{1}{\sqrt 2}(x^\mu - \im x^{\mu +4})
\end{equation}
with derivatives
\begin{equation}\label{3.16}
\dpar_{z^\mu}^{}=\frac{\dpar}{\dpar z^\mu}=\frac{1}{\sqrt 2}\,\bigl (\dpar_\mu -\im\dpar_{\mu +4}\bigr)\und 
\dpar_{\zb^{\bar \mu}}=\frac{\dpar}{\dpar\zb^{\bar \mu}}=\frac{1}{\sqrt 2}\,\bigl (\dpar_\mu +\im\dpar_{\mu +4}\bigr)\ .
\end{equation}
Note that $\dpar_\mu$ and $\dpar_{\mu +4}$ form a basis of the tangent space $\CV=\R^{6,2}$ to $T^*\R^{3,1}$ and the operator $\CJ=(\CJ_N^M)\in\,$End$(\CV)$ acts on \eqref{3.16} by formulae
\begin{equation}\label{3.17}
\CJ\left (\frac{\dpar}{\dpar z^\mu} \right )=\im\, \frac{\dpar}{\dpar z^\mu}  \und   
\CJ\left (\frac{\dpar}{\dpar \zb^{\bar\mu}} \right )=-\im\, \frac{\dpar}{\dpar \zb^{\bar\mu}}                
\end{equation}
which follow from formulae
\begin{equation}\label{3.18}
\CJ (\dpar_\mu )=\CJ_\mu^{\nu +4}\dpar_{\nu +4}=\dpar_{\mu +4}\und \CJ(\dpar_{\mu +4})=\CJ_{\mu +4}^\nu\dpar_\nu =-\dpar_\mu\ .
\end{equation}
In the basis $\dpar_{z^\mu}$, $\dpar_{\zb^{\bar\mu}}$, matrix $\CJ$ has the diagonal form $\CJ=\im\,\diag (\unit_4, -\unit_4)$.

\noindent {\bf Complex hyperbolic space.} Symplectic 2-form \eqref{2.1} and metric \eqref{2.7} in complex coordinates \eqref{3.15}  have the form 
\begin{equation}\label{3.19}
\omega_{\R^{6,2}}^{}=\frac{\im}{w^2}\,\eta_{\mu\bar\nu}\dd z^\mu\wedge \dd\zb^{\bar\nu}\und
g_{\R^{6,2}}^{}=\eta_{\mu\bar\nu}(\dd z^\mu\otimes \dd\zb^{\bar\nu} + 
\dd\zb^{\bar\nu}\otimes\dd z^\mu )\ .
\end{equation}
Thus, the complex structure $\CJ$ defines a pseudo-K\"ahler structure on the space $\R^{6,2}\cong\C^{3,1}$. Solution \eqref{3.5} for the KG oscillator in complex coordinates \eqref{3.15} has the form
\begin{equation}\label{3.20}
z^\mu(\tau ) = e^{\im\omega\tau} z^\mu\for z^\mu=z^\mu(0)=\frac{1}{\sqrt 2}(x^\mu - \im w^2 p^\mu )\ ,
\end{equation}
where the numbers $z^\mu$ parametrize initial data. The level surface in these coordinates is given by the equation
\begin{equation}\label{3.21}
\Ad S_7:\quad \frac{2}{w^4}\, \eta_{\mu\bar\nu}^{}\, z^\mu \zb^{\bar\nu} =-m^2\ ,
\end{equation}
where instead of $z^\mu (\tau)$ one can use $z^\mu =z^\mu (0)$ due to the independence of \eqref{3.21} from $\tau$.

The complex structure \eqref{3.14} can be associated with the vector field
\begin{equation}\label{3.22}
\CJ := \CJ_\mu^{\nu +4}x^\mu\dpar_{\nu +4} + \CJ^\nu_{\mu +4}x^{\mu +4}\dpar_{\nu }=x^\mu\dpar_{\mu +4}-
x^{\mu +4}\dpar_{\mu }
\end{equation}
and comparing \eqref{3.22} with \eqref{3.3} we see that $V_H=\omega\CJ$, i.e. $\CJ$ is the generator of the group U(1) introduced in \eqref{3.9}. In complex coordinates, the vector field \eqref{3.22} has the form
\begin{equation}\label{3.23}
\CJ = \im \left(z^\mu\,\frac{\dpar}{\dpar z^\mu} -\zb^{\bar\mu}\frac{\dpar}{\dpar \zb^{\bar\mu}}\right)\ .
\end{equation}
This Lie group defines the dynamics -- the rotation of a particle along a circle $S^1$ in the fibre  of bundle \eqref{3.10}, and the equations 
\begin{equation}\label{3.24}
\dot z^\mu =\omega\CJ (z^\mu)=\im\omega z^\mu\quad\Rightarrow\quad z^\mu (\tau)=e^{\omega\tau\CJ}z^\mu =e^{\im\omega\tau}z^\mu 
\end{equation}
are the infinitesimal form of the condition for the invariance of the level surface Ad$S_7$ with respect to the action of the group U(1). Solutions of the flow equations \eqref{3.24} define {\it orbits} of group U(1) in Ad$S_7$.

The space of orbits of U(1) in Ad$S_7$ can be identified with the complex hyperbolic space,
\begin{equation}\label{3.25}
H^3_\C =\Ad S_7/\sU(1)=\mbox{PU}(3, 1)/\sU(3){\cong}\sSU(3,1)/\rS(\sU(3){\times}\sU(1))\ ,
\end{equation}
for $p_0>0$ or with the complex conjugate space $\overline{H^3_\C}$ for $p^0<0$. Space ${H^3_\C}$ is simply connected manifold and can be identified with a complex 3-ball in $\C^3$ with a boundary $S^5=\dpar H_\C^3$ (see e.g. \cite{Ahmed, Diaz}). The variety $H^3_\C$ is a projectivization of the space $\C^{3,1}\cong\R^{6,2}$ and is covered by one patch, so that
\begin{equation}\label{3.26}
\C^{3,1}\ni (z^\mu(\tau))=(z^0(\tau), z^a(\tau)) \to \left(1,\frac{z^a(\tau)}{z^0(\tau)}\right)=\left(1,\frac{z^a(0)}{z^0(0)}\right)=:(1, y^a)\in H^3_\C\ .
\end{equation}
From \eqref{3.21} it follows that $2z^0\zb^{\bar 0}\ge m^2w^4$ and therefore the coordinates $y^a$ on $H_\C^3$ must satisfy the condition
\begin{equation}\label{3.27}
\delta_{a\bar b}\,y^a\yb^{\bar b}<1\ ,
\end{equation}
i.e. they parametrize the open 3-ball in $\C^3$. Also, $y^a$ do not depend on $\tau$ and parametrize the initial data of the KG oscillator. 

\section{Relativistic QM as gauge theory on $T^*\R^{3,1}$}

\noindent {\bf Geometric quantization}. More than 40 years ago, in the geometric quantization approach, it was shown that nonrelativistic quantum mechanics can be considered as a gauge theory of special type on phase spaces of classical particles \cite{Sour, Kost, Sni, Wood}. Namely, if phase space of a particle is a $2n$-dimensional manifold $(X, \omega_X)$ with a symplectic 2-form $\omega_X(=\dd\theta_X$ locally), then quantum mechanics is defined in terms of~\footnote{We use the natural units with $\hbar =c=1$.}
\begin{itemize}
\item a complex line bundle $L_\C^+$ over $X$ with structure group U(1)$_{\sf{v}}$, connection $\Av =\im\,\theta_X$ and curvature $\Fv =\dd\Av=\im\,\omega_X$,
\item wave functions $\psi\in\Gamma (X, L_\C^+)$ which are smooth sections of the bundle $L_\C^+\to X$,
\item polarization $\CT$ on $X$, defined as integrable Lagrangian subbundle of the complexified tangent bundle $T^\C X$ of $X$,
\item polarization condition $\dd_\CT^{}\psi =0$ meaning that the wave functions $\psi\in\Gamma (X,  L_\C^+)$ do not depend on coordinates on an integral manifold of the distribution $\CT\subset T^\C X$.
\end{itemize}
The abbreviation ``$\sf v$" and ``$\sf vac$"  here mean ``vacuum" since $\theta_{X}^{}$ and $\omega_{X}^{}$ have no sources and define the canonical symplectic structure on $X$.
Thus, nonrelativistic quantum mechanics is defined as a theory of complex scalar field $\psi$ on a manifold $(X, \omega_X)$, where $\psi$ is acted upon by covariant derivative $\nabla_A=\dd + \Av$ with $\Av = \im\,\theta_X$
defined on the bundle $L_\C^+$ over $X$, 
\begin{equation}\nonumber
{\rm QM}=\Bigl(L_\C^+\to X,\  \nabla_A,\ \psi\in\Gamma (X, L_\C^+),\ \CT\subset T^\C X,\ \dd^{}_\CT\psi =0\Bigr)\ .
\end{equation}
We emphasize that this data defines ``pure" quantum mechanics without interaction with electromagnetic and other physical fields. Hamiltonians are not involved in the above definition, i.e. it defines the ``kinematics" of QM. The field $\Av$ is fixed by the symplectic potential $\theta_X$ up to an automorphism of the bundle $L_\C^+$ and has no sources.

Fixing a polarization $\CT$ corresponds to the choice of an irreducible position, momentum, holomorphic or antiholomorphic representations for the canonical commutation relations (CCR) and they are all unitarily equivalent due to Stone-von~Neuman theorem. We show that after choosing a polarization $\CT$, the components of covariant derivative $\nabla_A$ coincide with the position and momentum operators or with the creation and annihilation operators in the Segal-Bargmann representation \cite{Segal, Bar, Hall}. Thus, specifying the fields $\Av$ and $\Fv = \dd\Av$ is equivalent to specifying basic observables and commutation relations for them.

Recall that if $E$ is a complex vector bundle (including line bundle) then there are three more complex vector bundles: the complex conjugate bundle $\bar E$, the dual bundle $\Ev$ and the dual of the complex conjugate bundle $\bar E^{\sf{v}}$. If $E$ is a Hermitian complex vector bundle then bundles $\bar E$ and $\Ev$ are isomorphic as well as bundles $E$ and $\bar E^{\sf{v}}$. Bundles $E$ and $\bar E$ are not isomorphic and in particular the $k$-th Chern class of $\bar E$ is given by $c_k(\bar E)=(-1)^k c_k(E)$ so that $c_1(\bar E)=- c_1(E)$. The bundle $L_\C^+$ introduced in quantum mechanics is a Hermitian complex line bundle \cite{Sour}-\cite{Wood} and therefore it is possible to introduce a non-isomorphic complex conjugate bundle $L_\C^- :=\overline{L_\C^+}$ \cite{Popov1, Popov2}. Sections $\Psi_\pm$ of these bundles $L_\C^\pm$ have a charge $\qv =\pm 1$ associated with the structure group $U(1)_{\sf{v}}$. This charge is called quantum charge in \cite{Popov1, Popov2}, it distinguishes particles ($\qv =1$) and antiparticles ($\qv =-1$). Abelian connection and curvature on bundles $L_\C^\pm$ have opposite signs: $A^\pm_{\sf{vac}}=\pm\im \theta_X$ and $F^\pm_{\sf{vac}}=\pm\im \omega_X$.

Having considered the geometry of phase spaces of free relativistic particles and KG oscillators, we move on to defining relativistic quantum mechanics (RQM) of spinless particles as a gauge theory on the phase space  $T^*\R^{3,1}$ with symplectic structure \eqref{2.1}-\eqref{2.4} and metric \eqref{2.7}-\eqref{2.9}. Namely, generalizing the nonrelativistic case, we define the RQM as a set of objects,
\begin{equation}\label{4.1}
\mbox{RQM}=\Bigl(L_\C^\pm ,\ \nabla_A,\ \Psi_\pm\in\Gamma(T^*\R^{3,1}, L_\C^\pm),\ \CT_\pm :\ \dd^{}_{\CT_\pm}\Psi_\pm =0\Bigr),
\end{equation}
where the phase space $T^*\R^{3,1}$ is equipped not only with a symplectic form but also with a pseudo-Euclidean metric of signature (6,2).

\noindent {\bf Complex spaces $V^\pm$.} To describe complex line bundles $L_\C^\pm$ from \eqref{4.1}, it is necessary to describe their fibres $V^\pm$. They are introduced as follows (see, for example, \cite{KobNom}). Consider a two-dimensional column
\begin{equation}\label{4.2}
\Psi =\begin{pmatrix}\psi^1\\ \psi^2\end{pmatrix}\in\C^2
\end{equation}
as a complex vector with coordinates $\psi^1, \psi^2\in\C$. Lie group SO(2)$_{\sf v}\cong\sU(1)_{\sf v}$ with generator
\begin{equation}\label{4.3}
J=\begin{pmatrix}0&-1\\ 1&0\end{pmatrix}
\end{equation}
acts on such columns. Operator $J$ has the following eigenvectors
\begin{equation}\label{4.4}
Jv_\pm = \pm\im v_\pm\ \ \Rightarrow\ \ v_\pm = \frac{1}{\sqrt 2}\,\begin{pmatrix}1\\ \mp\im\end{pmatrix},\ \ 
v_-=v_+^* ,\ \ v_\pm^\+ v^{}_\pm =1 ,\ \ v_\pm^\+ v^{}_\mp =0\ ,
\end{equation}
where ``$\ast$" means complex conjugation. Vectors $v^{}_\pm$ are the basis vectors of two orthogonal subspaces $V^{\pm}$ in $\C^2 = V^+\oplus V^- =\C\oplus\bar\C$. Any vector $\Psi$ from \eqref{4.2} can be decomposed into (1,0)- and (0,1)-parts:
\begin{equation}\label{4.5}
\Psi = \begin{pmatrix}\psi^1\\ \psi^2\end{pmatrix}=\Psi_+ + \Psi_- =\psi_+v_+ + \psi_-v_-\in\C\oplus\bar\C\with 
\psi_\pm = \frac{1}{\sqrt 2}\,(\psi^1\pm\im\psi^2)\ ,
\end{equation}
where $\Psi_\pm\in V^\pm$.

\noindent {\bf Quantum charge.} The Lie group U(1)$_{\sf v}$ acts on the space \eqref{4.5} of $\C^2$-vectors by multiplying on the left by matrices
\begin{equation}\label{4.6}
e^{\theta J}=\cos\theta + J\sin\theta = \begin{pmatrix}\cos\theta&-\sin\theta\\ \sin\theta&\cos\theta\end{pmatrix}\in\sSO(2)_{\sf v}\ .
\end{equation}
For subspaces $V^\pm$ in $\C^2=V^+\oplus V^-$ we obtain
\begin{equation}\label{4.7}
\Psi (\theta) = e^{\theta J}\Psi = \Psi_+(\theta )+\Psi_-(\theta ) = e^{\im\theta}\psi_+v_+ + e^{-\im\theta}\psi_-v_-\ ,
\end{equation}
and the action of the generator $\dpar_\theta$ of the group U(1)$_{\sf v}$ on $\Psi (\theta)$ has the form
\begin{equation}\label{4.8}
\dpar_\theta\Psi (\theta )=J\Psi (\theta) =\im\Psi_+(\theta )-\im\Psi_-(\theta )\ ,
\end{equation}
i.e. it is equivalent to the action of the generator $J$ from \eqref{4.3}. The group U(1)$_{\sf v}$ acting in the fibres $V^\pm$ of the bundles $L_\C^\pm$ can be associated with the charge $\qv$ as an eigenvalue of the operator $Q_{\sf v}$,
\begin{equation}\label{4.9}
Q_{\sf v}:= -\im J:\quad Q_{\sf v}\Psi_\pm = \qv\Psi_\pm =\pm\Psi_\pm \ .
\end{equation}
In \cite{Popov1, Popov2}, this charge was called quantum charge that distinguishes particles ($\qv =1$) and antiparticles ($\qv =-1$).

\noindent {\bf Complex line bundles $L_\C^\pm$.} Having complex spaces $V^\pm$, we can introduce complex line bundles $L_\C^\pm$ over the relativistic phase space $T^*\R^{3,1}$. In the case under consideration, these are direct products
\begin{equation}\label{4.10}
L_\C^\pm = T^*\R^{3,1}\times V^\pm
\end{equation}
with the base $T^*\R^{3,1}$ and fibres $V^\pm$. These two bundles are complex conjugate to each other and are associated with quantum charges $\qv =1$ and $\qv =-1$, respectively.

We introduce a Hermitian structure on the bundles $L_\C^\pm$ by equipping fibres $V^\pm$ with the Hermitian metric
\begin{equation}\label{4.11}
\langle \psi_\pm , \psi_\pm\rangle = \Psi_\pm^\+\Psi_\pm = \psi_\pm^*\psi_\pm\ .
\end{equation}
It is obviuos that the metric \eqref{4.11} is invariant under the action of the group U(1)$_{\sf v}$: $\psi_\pm\mapsto g_\pm\psi_\pm$ with $g_\pm =\exp (\pm\im\theta )\in\sU(1)_{\sf v}$.

\noindent {\bf Connection $\Av$}. Next, in the bundles $L_\C^\pm$ we should define connections
\begin{equation}\label{4.12}
A^\pm_{\sf vac} =\pm\im \theta_{\R^{6,2}}^{}=\pm\im\bigl (\sfrac12 p_\mu\dd x^\mu - \sfrac12 x^\mu\dd p_\mu\bigr )=\pm\frac{\im}{2w^2}\,\eta_{\mu\nu} \bigl (x^\mu\dd x^{\nu +4} -  x^{\mu +4}\dd x^\nu \bigr)\ ,
\end{equation}
where $\theta_{\R^{6,2}}^{}$ is the potential for the symplectic 2-form $\omega_{\R^{6,2}}^{}=\dd\theta_{\R^{6,2}}^{}$. Using the generator $J$ of group $\sU(1)_{\sf v}$, one can combine the bundles $L_\C^\pm$ into one $\C^2$-vector bundle
\begin{equation}\label{4.13}
L_{\C^2}=L_\C^+\oplus L_\C^-
\end{equation}
with connection and curvature of the form
\begin{equation}\label{4.14}
\Av =\theta_{\R^{6,2}}^{}J\und \Fv = \dd \Av =\omega_{\R^{6,2}}^{}J\ ,
\end{equation}
with eigenvalues $\pm \im$ of the operator $J$ on the subbundles $L_\C^\pm$.

Both $\Av$ and $\Fv$ take values in the Lie algebra $\ru(1)_{\sf v}=\,$Lie\,U(1)$_{\sf v}$. The abbreviation ``$\sf v$" and ``$\sf vac$" here mean ``vacuum" since $\theta_{\R^{6,2}}^{}$ and $\omega_{\R^{6,2}}^{}$ have no sources and define the canonical symplectic structure on $T^*\R^{3,1}$. Therefore, the background gauge fields $\Av$ and $\Fv$ define a vacuum of quantum mechanics. It is the constant components of the field $\Fv$ define the canonical commutation relations (CCR) and make the vacuum contribution to the calculation of all quantities. In fact, the field $\Av$ specifies interaction of particles with quantum charge $\qv\ne 0$ with vacuum and is associated with the potential energy of particles.

\noindent {\bf Curvature $\Fv$}. Using \eqref{4.12} and \eqref{4.14}, we can rewrite the connection $\Av$ in the form
\begin{equation}\label{4.15}
\Av = A_\mu J\dd x^\mu + A_{\mu +4}J\dd x^{\mu +4}
\end{equation}
with
\begin{equation}\label{4.16}
A_\mu =-\frac{1}{2w^2}\,\eta_{\mu\nu}x^{\nu +4}=\frac12\,p_\mu\und
A_{\mu +4}=\frac{1}{2w^2}\,\eta_{\mu\nu}x^{\nu}=:-\frac{1}{w^2}\,\eta_{\mu\nu}A^{\nu+4}\ .
\end{equation}
Accordingly, the covariant derivative $\nabla_A$ on $L_{\C^2}$ has components
\begin{equation}\label{4.17}
\nabla_\mu =\dpar_\mu + A_\mu J\und \nabla_{\mu +4}=\dpar_{\mu +4} + A_{\mu +4}J
\end{equation}
and in terms of coordinates and momenta we have
\begin{equation}\label{4.18}
\nabla_\mu =\dpar_\mu +\frac12\, p_\mu J ,\quad \nabla_{\mu +4}=-\frac{1}{w^2}\,\eta_{\mu\nu}\nabla^{\nu +4}:=-\frac{1}{w^2}\,\eta_{\mu\nu}\left(\frac{\dpar}{\dpar p_\nu}-\frac12\,x^\nu J\right)\ .
\end{equation}
Using \eqref{4.15}-\eqref{4.18} we obtain the following expressions for the curvature
\begin{equation}\label{4.19}
\Fv =\CF_{\mu\,\nu+4}\dd x^\mu\wedge \dd x^{\nu +4}\quad\Rightarrow\quad \CF_{\mu\,\nu+4}=F_{\mu\,\nu+4}J=
\frac{1}{w^2}\,\eta_{\mu\nu} J
\end{equation}
and 
\begin{equation}\label{4.20}
\CF_\mu^{~\nu +4} =[\nabla_\mu , \nabla^{\nu +4}]=\left[\dpar_\mu +\frac12\, p_\mu J ,\ \frac{\dpar}{\dpar p_\nu}-\frac12\,x^\nu J\right]=-\delta_\mu^\nu J\ .
\end{equation}
Recall that on bundles $L_\C^\pm$ the operator $J$ is replaced by $\pm\im$.

\noindent {\bf Polarizations $\CT_\pm$}. We have introduced bundles $L_\C^+$ and $L_\C^-$, sections of which $\Psi_+$ and $\Psi_-$ define particles ($\qv=1$) and antiparticles ($\qv=-1$). We have also introduced vacuum gauge field $\Av$. We will now discuss polarizations.

The spaces $\Gamma (L_\C^\pm)$ of sections $\Psi_\pm$ of bundles $L_\C^\pm$ are too large and they need to be narrowed down to spaces of irreducible representations of CCR by imposing conditions
\begin{equation}\label{4.21}
X_\pm^{}\Psi_\pm^{}=0\for \Psi_\pm^{}\in\Gamma (L_\C^\pm), \quad X_\pm^{}\in \Gamma (\CT_\pm), \quad \CT_\pm\subset T^\C\R^{6,2}\ ,
\end{equation}
where $X_\pm^{}$ are vector fields from the subbundles $\CT^{}_\pm$ of the complexified tangent bundle of the phase space $T^*\R^{3,1}$. Conditions \eqref{4.21} are usually imposed in combination with one or another automorphism of the bundle $L_\C^\pm$. 

Usually, one of two real polarizations $\CT_+=\CT_-=\CT$ is considered -- either the independence of $\Psi_\pm$ from momenta, or their independence from coordinates. We will consider only the position representation specified by the conditions
\begin{equation}\label{4.22}
\frac{\dpar\Psi_\pm^{}}{\dpar p_\mu}=0\quad\Rightarrow\quad\Psi_\pm^{}=\Psi_\pm^{}(x^\mu , \tau ).
\end{equation}
To define complex polarizations $\CT^{}_\pm$, one has to define two conjugate complex structures $\CJ_\pm=\pm\CJ$ and impose holomorphicity conditions on $\Psi_\pm^{}$ with respect to these complex structures $\CJ_\pm$. We will first consider polarization \eqref{4.22} and the associated operators $\hat p_\mu$ and $\xh^\mu$, and defer consideration of complex polarizations to the next section.

\noindent {\bf Configuration space representation}. Let us consider constraints \eqref{4.22} on $\Psi_\pm$. Note that the vector fields $\dpar/\dpar p_\mu$ do not commute with the covariant derivatives \eqref{4.18}, which is unacceptable. However, connection $\Av$ in \eqref{4.16} can be transformed using the action of the group $\CG$ of unitary automorphisms of the bundle $L_{\C^2}$,
\begin{equation}\label{4.23}
\CG = C^\infty (T^*\R^{3,1}, \sU(1)_{\sf v}),
\end{equation}
with elements $g=\exp (\al (x,p)J)$, where $\al (x,p)$ is a real function on $T^*\R^{3,1}$. If we choose $\al =-\sfrac12\,p_\mu x^\mu$, we obtain
\begin{equation}\label{4.24}
A_\mu^\al =A_\mu + g^{-1}\dpar_\mu g=0\ ,\quad A^\al_{\mu +4}=A^{}_{\mu +4}+g^{-1}\dpar^{}_{\mu +4}g=\frac{1}{w^2}\,\eta_{\mu\nu}x^\nu\ \Rightarrow\ A^{\al , \mu +4}=-x^\mu
\end{equation}
\begin{equation}\label{4.25}
\Rightarrow\quad\nabla^\al_{\dpar/\dpar x^\mu}=\dpar_\mu\und \nabla^\al_{\dpar/\dpar p_\mu}=\frac{\dpar}{\dpar p_\mu}- x^\mu J\ .
\end{equation}
Now the covariant derivatives \eqref{4.25} commute with the derivatives in \eqref{4.22} and we can introduce the operators
\begin{equation}\label{4.26}
\xh^\mu := J\nabla^\al_{\dpar/\dpar p_\mu}=x^\mu +J\frac{\dpar}{\dpar p_\mu}\und 
\hat p_\mu :=-\im\nabla^\al_{\dpar/\dpar x^\mu}=-\im\frac{\dpar}{\dpar x^\mu}\ .
\end{equation}
These observables have canonical commutation relations
\begin{equation}\label{4.27}
[\hat p_\mu , \xh^\nu ]=-\im J\CF_\mu^{~\nu +4}=-\im\delta_\mu^\nu\ ,
\end{equation}
where $\CF_\mu^{~\nu +4}$ are the components of the vacuum field $\Fv$ given in \eqref{4.20}. Note that the second term in the operators $\xh^\mu$ disappears on functions satisfying \eqref{4.22} and on such functions the operators $\xh^\mu$ and $\hat p_\mu$ acquire the standard form. In this case, the operator $\xh^0$ is a quantum time operator which is an observable of the same type as $\xh^a$ due to the fact that the components $\hat p_\mu$ of the momentum are not constrained by the energy-momentum relation \eqref{2.16}. Dynamics is defined through the proper time $\tau$ so that $x^0$ can be a function of $\tau$, similar to the position variables $x^a$ of a particle.

\section{Relativistic QM in holomorphic representation}

\noindent {\bf Complex coordinates $z^\mu$}. 
We have described all the objects from \eqref{4.1} necessary to define  relativistic quantum mechanics, including real polarization \eqref{4.22}. Here we will describe two pseudo-K\"ahler  polarizations.

 Recall  that on space $T^*\R^{3,1}$ we introduced the symplectic form \eqref{2.1} and metric \eqref{2.7} of signature (6,2). These two structures make it possible to fix the complex structure \eqref{3.13}-\eqref{3.15} invariant under transformations SU(3,1) preserving the pseudo-K\"ahler structure on phase space $\C^{3,1}\cong T^*\R^{3,1}$ that combines these metric and symplectic form. In complex coordinates \eqref{3.15} they have the form \eqref{3.19} and can be combined into the pseudo-Hermitian metric
\begin{equation}\label{5.1}
h_{\C^{3,1}}^{} = g_{\R^{6,2}}^{}-\im\,w^2\,\omega_{\R^{6,2}}^{}=2\eta_{\mu\bar\nu}\dd z^\mu\otimes \dd\zb^{\bar\nu}
\end{equation}
on the space $\C^{3,1}$.

\noindent {\bf Covariant derivatives $\nabla^{}_{z^\mu_\pm}$.} One-form of connection \eqref{4.15} on the bundle $L_{\C^2}^{}=L_\C^+\oplus L_\C^-$ has the form
\begin{equation}\label{5.2}
A_{\sf vac} =\frac{1}{2w^2}\bigl (x^\mu\dd x^{\nu +4}- x^{\mu+4}\dd x^\nu\bigr )J=\frac{m\omega}{2}\,\eta_{\mu\bar\nu} \bigl (\zb^\nb\dd z^{\mu} -  z^{\mu}\dd \zb^\nb \bigr) J\ .
\end{equation}
Accordingly, for covariant derivatives we obtain expressions
\begin{equation}\label{5.3}
\nabla_{z^\mu}^{}=\dpar_{z^\mu}^{} +  \frac{1}{2w^2}\eta_{\mu\bar\nu}\zb^{\bar \nu}Q_{\sf v}\ ,\quad
\nabla_{\zb^{\bar \mu}}^{}=\dpar_{\zb^{\bar \mu}} -\frac{1}{2w^2}\eta_{\bar\mu\nu}^{}z^{\nu}Q_{\sf v}\ ,
\end{equation}
and expressions for derivatives $\nabla^\pm$ on subbundles $L_\C^\pm$ are obtained by replacing $Q_{\sf v}\to\pm 1$. We emphasize that the covariant derivatives acting on $\Psi_\pm\in L_\C^\pm$ have different forms since $\Psi_\pm$ have opposite sign of the quantum charge $\qv =\pm 1$.

For a uniform description of covariant derivatives on the bundles $L_\C^+$ and  $L_\C^-$, we introduce the notation
\begin{equation}\label{5.4}
z_+^\mu := z^\mu\und z_-^\mu := \zb^\mb =(z_+^\mu )^*\ .
\end{equation}
In these coordinates, the covariant derivatives on $L_\C^\pm$ take the same form
\begin{equation}\label{5.5}
\nabla_{z^\mu_\pm}=\dpar_{z^\mu_\pm} +  \frac{1}{2w^2}\eta_{\mu\bar\nu}\zb_\pm^{\bar \nu}\ ,\quad
\nabla_{\zb_\pm^{\bar \mu}}=\dpar_{\zb_\pm^{\bar \mu}} -
\frac{1}{2w^2}\eta_{\bar\mu\nu}z_\pm^{\nu}\ .
\end{equation}

\noindent {\bf Dolbeault operators.} We want to introduce holomorphic sections $\psi_\pm$ of bundles $L_\C^\pm$, i.e. those that satisfy the conditions
\begin{equation}\label{5.6}
\frac{\dpar\psi_\pm^{}}{\dpar\zb_\pm^\mb} =0\ \ \Rightarrow\ \ \psi_+=\psi_+(z_+^\mu )=\psi_+(z^\mu )\und \psi_-=\psi_-(z_-^\mu )=\psi_-(\zb^\mb )\ .
\end{equation}
Note that the vector fields $\dpar/\dpar{\zb_\pm^\mb}$ do not commute with the covariant derivatives \eqref{5.5} and therefore before imposing conditions \eqref{5.6} we must find automorphisms of the bundles $L_\C^\pm$ that reduce operators \eqref{5.5} to a form compatible with conditions \eqref{5.6}. To do this, we introduce the Dolbeault operators
\begin{equation}\label{5.7}
\bar\dpar^{}_{L_\C^\pm} = \dd\zb_\pm^{\bar\mu}\, \left (\frac{\dpar}{\dpar \zb_\pm^{\bar\mu}}+\frac{1}{2w^2}\eta_{\bar\mu\nu}z^\nu_\pm\right )
\end{equation}
and impose the conditions
\begin{equation}\label{5.8}
\bar\dpar^{}_{L_\C^\pm}\Psi_\pm =0
\end{equation}
on sections $\Psi_\pm$ of Hermitian complex line bundles $L_\C^\pm$. 

Solutions of \eqref{5.8} are vectors
\begin{equation}\label{5.9}
\Psi_\pm =\psi_\pm (z^\mu_\pm, \tau )v_\pm^c\with v_\pm^c=\psi_0(z,\zb ) v_\pm\ ,
\end{equation}
where 
\begin{equation}\label{5.10}
\psi_0(z,\zb )=\exp\Bigl(-\frac{1}{2w^2}\eta_{\mu\bar\nu}z^\mu_\pm \zb_\pm^{\bar \nu}\Bigr )
=\exp\Bigl(-\frac{1}{2w^2}\eta_{\mu\bar\nu}z^\mu \zb^{\bar \nu}\Bigr )\ .
\end{equation}
Inner product of $\C^2$-vectors \eqref{5.9} has the form
\begin{equation}\label{5.11}
\Psi_\pm^\+\Psi_\pm =\psi_\pm^*(z_\pm , \tau )\psi_\pm(z_\pm , \tau )\exp\Bigl(-\frac{1}{w^2}\eta_{\mu\bar\nu}z^\mu_\pm \zb_\pm^{\bar \nu}\Bigr )\ ,
\end{equation}
as it should be in the Segal-Bargmann representation \cite{Segal, Bar, Hall}. However, the pseudo-Hermitian metric in the exponentials \eqref{5.10} and \eqref{5.11} is not positive definite, so the Fock spaces $\CF^\pm$ of functions of the form \eqref{5.9} are not Hilbert spaces. This is the difference from the non-relativistic QM, where instead of $\eta_{\mu\nu}$ in \eqref{5.11} there is a Euclidean metric. It should be understood that functions \eqref{5.9}-\eqref{5.11} are defined on the space $\C^{3,1}\cong (T^*\R^{3,1}, \CJ)$, and not on the space of initial data of dynamic equations (covariant phase space), which was introduced in \eqref{3.25}. In the next section we will show that the dynamics fixes in the spaces $\CF^\pm$ well-defined Hilbert subspaces $\CH^\pm\subset\CF^\pm$ of functions on the covariant phase space.

\noindent {\bf Creation and annihilation operators}. The geometric meaning of  sections \eqref{5.9} of the bundles $L_\C^\pm$ is as follows. Basis vectors $v_\pm$ in fibres of these bundles define the Hermitian metrics \eqref{4.11}. These vectors $v_\pm^{}$ are Hermitian bases of Hermitian bundles and the covariant derivatives \eqref{4.17}, \eqref{4.18} and \eqref{5.3}, \eqref{5.5} with u(1)$_{\sf v}$-valued connections preserve this Hermitian structure. On the other hand, vectors $v_\pm^c$ in \eqref{5.9} define in $L_\C^\pm$ a complex basis associated with the principal bundle $P(T^*\R^{3,1}, \sGL(1, \C)_{\sf v})$ having the structure group $\sGL(1, \C)_{\sf v}=\C^*\supset\sU(1)_{\sf v}$ \cite{KobNom}. The function \eqref{5.10} in \eqref{5.9}-\eqref{5.11} is an element of the group $\sGL(1, \C)_{\sf v}$ that defines a map of Hermitian bases  $v_\pm$ into holomorphic bases $v_\pm^c$ along which the holomorphic (w.r.t. complex structures $\pm\CJ$) sections \eqref{5.9} of the bundles $L_\C^\pm$ are decomposed.  Thus, the function $\psi_0$ from \eqref{5.10} is an element of the automorphism group Aut($L_\C^\pm$) that transforms the Hermitian structure into a holomorphic one.

Using function $\psi_0\in\sGL(1,\C)_{\sf v}$  as an automorphism of $L_\C^\pm$, we obtain
\begin{equation}\label{5.12}
\begin{split}
\nabla_{z^\mu_\pm}^{\psi_0}=\dpar_{z^\mu_\pm} +  A_{z^\mu_\pm} + \psi_0^{-1}\dpar_{z^\mu_\pm}\psi_0=
\frac{\dpar}{\dpar{z^\mu_\pm}}\ ,\\[2pt]
\nabla_{\zb_\pm^{\bar \mu}}^{\psi_0}=\dpar_{\zb_\pm^{\bar \mu}} + A_{\zb_\pm^{\bar \mu}}+ \psi_0^{-1}\dpar_{\zb^{\bar\mu}_\pm}\psi_0=\frac{\dpar}{\dpar{\zb^{\bar\mu}_\pm}} -\frac{1}{w^2}\eta_{\bar\mu\nu}z_\pm^{\nu}\ .
\end{split}
\end{equation}
Under this automorphism the Dolbeault operators \eqref{5.7} transform into ordinary $\bar\dpar$-operators
\begin{equation}\label{5.13}
\bar\dpar^{\psi_0}_{L_\C^\pm}=\dd\zb_\pm^{\bar\mu}\,\frac{\dpar}{\dpar{\zb^{\bar\mu}_\pm}}
\end{equation}
with partial derivatives $\dpar_{\zb^{\bar\mu}_\pm}$. They commute with transformed covariant derivatives \eqref{5.12}. 

It is easy to verify that on sections \eqref{5.9} the operators of covariant derivatives \eqref{5.5} have the form
\begin{equation}\label{5.14}
\nabla_{z_\pm^\mu}\Psi_\pm = \Bigl(\dpar_{z_\pm^\mu}\psi_{\pm}\Bigr)v_\pm^c\und \nabla_{\zb_\pm^{\bar \mu}}\Psi_\pm =-\Bigl(\frac{1}{w^2}\eta_{\bar\mu\nu} z_\pm^\nu\psi_\pm\Bigr)v_\pm^c
\end{equation}
i.e. they are reduced to \eqref{5.12}. It follows from this that the annihilation and creation operators for $\Psi_\pm$ from  \eqref{5.9} have the form
\begin{equation}\label{5.15}
a_{\mu\pm}^{}=w\nabla_{z_\pm^\mu}^{}\und a_{\bar\mu\pm}^\+=-w\nabla_{\zb_\pm^{\bar \mu}}^{}\with [a_{\mu\pm}^{}, a^\+_{\bar\nu\pm}]=\eta_{\mu\bar\nu}^{}\ .
\end{equation}
After transformations \eqref{5.12}, they take the usual form
\begin{equation}\label{5.16}
a_{\mu\pm}^{}=\dpar_{\tilde z_\pm^\mu}^{}\und a_{\mu\pm}^{\+}=\eta_{\bar\mu\nu}\tilde z_\pm^\nu\ \ \mbox{for}\ \    \tilde z_\pm^\mu = \frac{z_\pm^\mu}{w} \ ,
\end{equation}
when acting on holomorphic functions $\psi_\pm (z_\pm^\mu , \tau)$ of $z_\pm^\mu$ (Segal-Bargmann representation \cite{Segal, Bar, Hall}). Thus, the creation and annihilation operators are defined by covariant derivatives \eqref{5.5}, \eqref{5.15} when acting on $\Psi_\pm\in \CF^\pm$ and by operators \eqref{5.12}, \eqref{5.16} when acting on functions $\psi_\pm^{}(z_\pm^{} , \tau)$.

Note again that the Fock spaces $\CF^\pm$ contain negative-norm states arising from the $\eta_{0\bar 0}^{}$ component of the pseudo-Hermitian metric \eqref{5.1} on the ``kinematic" phase space $\C^{3,1}$. However, $\CF^\pm$ are not spaces of physical states. Note that the space of initial data of motion of any relativistic particle -- the covariant phase space -- is a six-dimensional manifold embedded in $\C^{3,1}$ with an induced metric of Euclidean signature. The function spaces of this six-dimensional manifold will be Hilbert spaces $\CH^\pm$ embedded in $\CF^\pm$ as subspaces. In what follows we describe the spaces $\CH^\pm$ for Klein-Gordon oscillators.

\section{Quantum Klein-Gordon oscillator}

\noindent {\bf Particles and antiparticles}. In Section 4, we described relativistic quantum mechanics of spinless particles as a gauge theory \eqref{4.1} of a special type with a gauge field $\Av\in\ru(1)_{\sf v}$ characterizing the vacuum and covariant derivatives $\nabla_A$ acting on polarized sections $\Psi_+$ and $\Psi_-$ of the complex conjugate line bundles $L_\C^+$ and $L_\C^-$. The curvature of this connection (the strength of the gauge field) is
\begin{equation}\label{6.1}
\Fv = \dd\Av =\omega_{\R^{6,2}}^{}J=\left\{\begin{array}{r}\im\omega_{\R^{6,2}}^{}\ \mbox{on}\ L_\C^+\\
-\im\omega_{\R^{6,2}}^{}\ \mbox{on}\ L_\C^-\end{array}\right.
\end{equation}
where  $\omega_{\R^{6,2}}^{}$ is a symplectic form on the phase space $T^*{\R^{3,1}}$.
Associated with these bundles are the charges $\qv=1$ and $\qv=-1$ so that $\Psi_+$ describes particles and $\Psi_-$ describes antiparticles. 

The Heisenberg algebra is defined by covariant derivatives $\nabla_A^\pm$ acting on sections of bundles $L_\C^\pm$. If we impose the polarization condition \eqref{4.21} on $\Psi_\pm$, we obtain irreducible representation of CCR on the space of such functions. In Section 4 we described real polarizations using the example of position representation \eqref{4.22}-\eqref{4.27}.
In Section 5, we described complex polarizations -- they are the ones associated with the creation and annihilation operators that are important in describing oscillators in QM, RQM and QFT. The introduction of conjugate complex structures $\CJ_+=\CJ$ and $\CJ_-=-\CJ$ on the relativistic phase space $T^*\R^{3,1}$ allows one to clearly see the difference between particles $\Psi_+$ and antiparticles $\Psi_-$ as holomorphic and antiholomorphic objects. In this section we will consider the interaction of these particles $\Psi_\pm$ with the fundamental field $\Av$ which leads to the Klein-Gordon oscillator equation.

\noindent {\bf Covariant Laplacian}. In position representation we have sections $\Psi_\pm = \psi_\pm (x^\mu , \tau)v_\pm\in L_\C^\pm$ and covariant derivatives \eqref{4.25} acting on $\Psi_\pm$.  Using metric tensor \eqref{2.7}-\eqref{2.9} on $T^*\R^{3,1}$, we introduce a covariant Laplacian acting on $\Psi_\pm$,
\begin{equation}\label{6.2}
\Delta_2=\eta^{\mu\nu}\nabla^\al_\mu\nabla^\al_\nu + \frac{1}{w^4}\,\eta_{\mu\nu}\nabla^\al_{\dpar/\dpar p_\mu}\nabla^\al_{\dpar/\dpar p_\nu}=
\eta^{\mu\nu}\dpar_\mu\dpar_\nu - m^2\omega^2\,\eta_{\mu\nu}x^\mu x^\nu\ ,
\end{equation}
where we have omitted derivatives with respect to momenta due to \eqref{4.22}. From \eqref{6.2} we see that the part of $\Delta_2$ along $x$-space defines the kinetic energy of particle, and the part of $\Delta_2$ along $p$-space specifies the potential energy. In fact, this potential energy arises from the interaction of particle with the vacuum field $\Av$. Potential energy of interaction with gauge field $A_\mu\dd x^\mu$ given in Minkowski space arises in a similar way through the extension of derivatives $\dpar_\mu\to \dpar_\mu +A_\mu$. 

We want to describe the quantum Klein-Gordon oscillator. To do this, we should introduce an operator version of the momentum map from \eqref{3.7},
\begin{equation}\label{6.3}
\hat\mu_H^{}=\frac1m\,\Delta_2\ ,
\end{equation}
which is the same for particles and antiparticles in position representation. Since $\Psi_+$ and $\Psi_-$ are sections of complex conjugate bundles, their evolution in $\tau$ is conjugate,
\begin{equation}\label{6.4}
\pm\im\dpar_\tau\Psi_\pm=\hat\mu_H^{}\Psi_\pm \ ,
\end{equation}
\begin{equation}\label{6.5}
\Psi_\pm =e^{\mp\im m\tau}\Phi_\pm (x^\mu )v_\pm\ \Rightarrow\ \hat\mu_H^{}\Phi_\pm =m\Phi_\pm\ \Rightarrow\ \bigl(\eta^{\mu\nu} \dpar_\mu\dpar_\nu -m^2- 
 m^2w^2\,\eta_{\mu\nu}x^\mu x^\nu  \bigr)\Phi =0  \ ,
\end{equation}
with
\begin{equation}\label{6.6}
\Phi = \Phi_+ +\Phi_- = \phi_+(x^\mu )v_+ + \phi_-(x^\mu )v_-\ .
\end{equation}
Here $\Phi_\pm$ are sections of the bundles $L_\C^\pm$, and $\Phi$ is a section of the bundle $L_{\C^2}$. 
Note that mass is introduced through the operator $\dpar_\tau$. 
Equation \eqref{6.5} is a ``quantum" version of the equation \eqref{3.6} for Ad$S_7$ level set and this is the Klein-Gordon oscillator equation. However, from the point of view of differential geometry, we have a bundle $L_{\C^2}$ with connection $\Av$, a covariant Laplacian on it, and a natural standard equation \eqref{6.5} on sections of this bundle. 
The last term in the KG   equation \eqref{6.5} arises from the interaction of fields $\Phi\in L_{\C^2}$ with the vacuum field $\Av$ present in the covariant derivatives \eqref{4.18}. Quantization consists only in replacing the partial derivatives $\dpar_M^{}$ on the phase space $T^*\R^{3,1}$ with covariant derivatives $\nabla_M$ acting no longer on functions $\phi$ on $T^*\R^{3,1}$, but on sections $\Phi$ of the bundle $L_{\C^2}$ over $T^*\R^{3,1}$, and that's all.

In the limit $\omega\to 0$, the interaction with the field $\Av$ disappears in \eqref{6.5} and we obtain the Klein-Gordon equation for ``free" particles $\Psi_+$ and $\Psi_-$ corresponding to the level surface $H^3\times \R^{3,1}$ from \eqref{2.16}. Therefore, solutions of the KG equation for free particles contain excess degrees of freedom. To eliminate them, one should move from 7-dimensional manifold $H^3\times \R^{3,1}$ to 6-dimensional manifold $T^*H^3$, which at the classical level is given by an additional equation $p_\mu x^\mu =0$ on the initial data, resolving which leads to the cotangent bundle $T^*H^3$. Accordingly, in relativistic QM a quantum version of this constraint must be added. We will consider this in more detail elsewhere.

\noindent {\bf Segal-Bargmann representation}. In the SB representation, $\Psi_\pm\in\CF^\pm$ are holomorphic sections of the bundles $L_\C^\pm$ and the spaces $\CF^\pm$ form irreducible representations of CCR with creation and annihilation operators \eqref{5.12}-\eqref{5.16}. These representations are not unitary. The quantum Klein-Gordon oscillator in this representation is again described by equation \eqref{6.4}, but now we have
\begin{equation}\label{6.7}
\hat\mu_H=\sfrac1m\,\Delta_2 = \sfrac1m\,g^{MN}\nabla_M\nabla_N=
\sfrac1m\,\eta^{\mu\nb}\bigl (\nabla_{z^\mu}^{}\nabla_{\zb^\nb}^{}+
\nabla_{\zb^\nb}^{}\nabla_{z^\mu}^{}\bigr)\ .
\end{equation}
Note that equations \eqref{6.4} can be combined into one equation,
\begin{equation}\label{6.8}
J\dpar_\tau\Psi =\sfrac1m\,\Delta_2\Psi\quad\Leftrightarrow\quad\pm\im\dpar_\tau\Psi_\pm=\sfrac1m\,\Delta_2\Psi_\pm\quad\mbox{for}\quad\Psi = \Psi_++\Psi_-\ ,
\end{equation}
where $\Delta_2$ is given in \eqref{6.7}.

It is easy to show that from \eqref{6.8} follow two continuity equations,
\begin{equation}\label{6.9}
\pm\dpar_\tau^{}\rho_\pm^{}+\nabla_M^{}j_\pm^M=\pm
\dpar_\tau^{}\rho_\pm^{}+\nabla_{z^\mu}^{}j_\pm^\mu +\nabla_{\zb^\mb}^{}j_\pm^\mb =0\ ,
\end{equation}
where
\begin{equation}\label{6.10}
\rho_\pm^{}:=\Psi_\pm^\+\Psi_\pm\ ,\quad j_\pm^M:=
\frac{\im}{m}\,g^{MN}\Bigl(\Psi_\pm^\+\nabla_N\Psi_\pm - (\nabla_N\Psi_\pm^\+)\Psi_\pm\Bigr)\ ,
\end{equation}
\begin{equation}\nonumber
j^\mu_\pm :=\frac{\im}{m}\,\eta^{\mu\nb}\Bigl(\Psi_\pm^\+\nabla_{\zb^\nb}^{}\Psi_\pm - 
(\nabla_{\zb^\nb}^{}\Psi_\pm^\+)\Psi_\pm\Bigr)\und
j^\mb_\pm :=\frac{\im}{m}\,\eta^{\mb\nu}\Bigl(\Psi_\pm^\+\nabla_{z^\nu}^{}\Psi_\pm - 
(\nabla_{z^\nu}^{}\Psi_\pm^\+)\Psi_\pm\Bigr)\ .
\end{equation}
Here $\pm\rho_\pm$ are densities of quantum charges and functions $\rho_\pm$ can be associated with probability densities in space-time. Note that in the limit of free particle $j_\pm^M$ is reduced to the expression
\begin{equation}\label{6.11}
j_{\mu\,\pm}^{}=\sfrac{\im}{m}\bigl(\Psi_\pm^\+\dpar_\mu\Psi_\pm - (\dpar_\mu\Psi_\pm^\+)\Psi_\pm\bigr)\ ,
\end{equation}
where $j_{0\,\pm}^{}\sim\pm\rho_\pm^{}$ is the quantum charge density, not the probability density.

Substituting into \eqref{6.8} the dependence on $\tau$ in the form
\begin{equation}\label{6.12}
\Psi_\pm =e^{\mp\im E_0\tau}\Phi_\pm(x,p)\ ,
\end{equation}
we obtain the equation of quantum relativistic oscillator
\begin{equation}\label{6.13}
(\Delta_2-m E_0 )\Phi_\pm =0\ .
\end{equation}
For the Klein-Gordon oscillator, the energy $E_0$ is not arbitrary, but fixed so that $E_0 =m$ as in \eqref{6.5} and from \eqref{6.9}-\eqref{6.13} we obtain the Klein-Gordon oscillator equation
\begin{equation}\label{6.14}
(\Delta_2-m^2)\Phi =0\quad\mbox{for}\quad\Phi = \Phi_+ + \Phi_-\ ,
\end{equation}
where $\Delta_2$ has the form \eqref{6.7}.

\noindent {\bf Solutions.} In the Introduction we already noted that in position representation the Klein-Gordon oscillator equation \eqref{6.2}-\eqref{6.6} does not have a good solution space -- either the unitary or the Lorentz covariance of the model is violated (see e.g. \cite{Bars, Bedic}). As was shown by Dirac \cite{Dirac}, on the space of solutions with the Euclidean ground state \eqref{1.3}, Lorentz covariance can be realized implicitly, but for this it is necessary to use negative powers of the operator $\xi^0$ from \eqref{1.5}. The reason for this is that the space of initial data of the KG oscillator is a coset space $Z_6=\Ad S_7/\sU(1)=\sU(3,1)/\sU(3)\times\sU(1)$ that does not have the structure of a cotangent bundle necessary for defining the coordinate or momentum representation. In fact, the space $(Z_6, \CJ)$ is a K\"ahler-Einstein manifold and it can be identified with the unit complex 3-ball in $\C^3$. Because of this homogeneous K\"ahler structure on $Z_6$, the Segal-Bargmann representation is well defined but the position representation is not.

Thus we seek solutions in the complex SB representations. It is easy to show that after substituting $\Psi_\pm =\psi_\pm (z_\pm )v_\pm^c$ into equations \eqref{6.14}, they are reduced to the equations
\begin{equation}\label{6.15}
\begin{split}
(\Delta_2 - m^2)(\Psi_++\Psi_-)=(\Delta_2 - m^2)(\psi_+v_+^c +\psi_-v_-^c)\\[1pt]
=-\frac{2}{w^2}\,\Bigl[(z_+^\mu\dpar_{z_+^\mu}+N+2)\psi_+v_+ + (z_-^\mu\dpar_{z_-^\mu}+N+2)\psi_-v_-\Bigr]\psi_0=0
\end{split}
\end{equation}
\begin{equation}\label{6.16}
\Rightarrow \bigl(z_\pm^\mu\dpar_{z_\pm^\mu}+N+2)\psi_\pm (z_\pm)=0\ ,
\end{equation}
where
\begin{equation}\label{6.17}
N:=\frac{m^2w^2}{2}=\frac{m}{2\omega}>0\ .
\end{equation}
We will consider $N$ as an interger fixed by the choice of parameter $\omega$.

Thus, we have shown that when using the pseudo-Hermitian structure \eqref{5.1} on the phase space $T^*\R^{3,1}$, equation \eqref{1.4} used by Dirac is replaced by equations \eqref{6.16} for holomorphic (particles) and antiholomorphic (antiparticles) functions on the space $\C^{3,1}=(T^*\R^{3,1}, \CJ)$. In \eqref{6.16} we have $z_+^\mu=z^\mu\in\C^{3,1}_+=\C^{3,1}$ and $z_-^\mu=\overline{z^\mu}\in\C^{3,1}_-=\overline{\C^{3,1}}$ and these are ordinary coordinates, not operators as in \eqref{1.4}.

General solutions of equations \eqref{6.16} have the form
\begin{equation}\label{6.18}
\psi_\pm =\biggl(\frac{1}{\sqrt 2\omega z_\pm^0}\biggr)^{N+2}f_\pm(y^1_\pm , y^2_\pm , y^3_\pm )\ ,
\end{equation}
where $f_\pm$ are arbitrary holomorphic functions of complex coordinates $y^a_\pm$ defined on unit 3-balls $B_\pm^3$ in $\C^3$,
\begin{equation}\label{6.19}
B_\pm^3=(Z_6, \pm\CJ)=\Bigl\{y_\pm^a =\frac{z^a_\pm}{z^0_\pm}\mid \delta_{a\bar b}\,y_\pm^a\yb_\pm^\bb <1\Bigr\}\ .
\end{equation}
Conditions \eqref{3.27} and \eqref{6.19} defining 3-balls $B_\pm^3$ with conjugate complex structure $(y_-^a=\overline{y_+^a})$ follow from the energy constraint \eqref{3.21}. We emphasize that constraints \eqref{6.19} do not limit the absolute value of coordinates and momenta since $y_\pm^a$ are the ratio of coordinates $z_\pm^\mu$.

\noindent {\bf Inner product.} Unconstrained sections $\Psi_\pm\in\CF^\pm$ of bundles $L_\C^\pm$ have the form 
\eqref{5.9} with an inner product \eqref{5.11}. However, for the quantum Klein-Gordon oscillator, $\Psi_\pm$ are not arbitrary, but satisfy two constraints. First, the coordinates $z_\pm^\mu$ satisfy the fixed energy equation \eqref{3.21} of the form
\begin{equation}\label{6.20}
-\frac{1}{w^2}\, \eta_{\mu\nb}z^\mu \zb^\nb = \frac{m^2w^2}{2}=N\ , 
\end{equation}
due to which the function $\psi_0$ from \eqref{5.10} becomes constant,
\begin{equation}\label{6.21}
\psi_0=\exp (\sfrac12\,N)\quad\Rightarrow\quad\psi_0^2=e^N\ ,
\end{equation}
in the inner product \eqref{5.11}. Second, $\psi_\pm(z_\pm)$ from \eqref{5.9} must satisfy equations \eqref{6.16}, the general solutions of which are given in \eqref{6.18}.

From \eqref{5.11}, \eqref{6.18} and \eqref{6.21} it follows that
\begin{equation}\label{6.22}
\Psi_\pm^\+\Psi_\pm^{}= e^Nf_\pm^*f_\pm^{}\,\mu_{N+2}^{}\ ,
\end{equation}
where
\begin{equation}\label{6.23}
\mu_{N+2}^{}:=\biggl(\frac{1}{ 2\omega^2 z_\pm^0\zb_\pm^{\bar 0}}\biggr)^{N+2}=
(1-\delta_{a\bb}\,y_\pm^a\yb_\pm^\bb)^{N+2}
\end{equation}
is a weight function. For two different solutions $\Psi$ and $\hat\Psi$ of equations \eqref{6.16} the inner product is defined as
\begin{equation}\label{6.24}
\langle\Psi_\pm^{} , \hat\Psi_\pm^{}\rangle := \int_{B_\pm^3}\Psi_\pm^\+\hat\Psi_\pm^{}\dd V = e^N\int_{B_\pm^3}f_\pm^*\hat f_\pm^{}\,\mu_{N+2}^{}\,\dd V\ ,
\end{equation}
where $\dd V=\im\vk^2\dd y^1\wedge\dd y^2\wedge\dd y^3\wedge\dd\yb^{\bar 1}\wedge\dd\yb^{\bar 2}\wedge\dd\yb^{\bar 3}$ and usually $\vk^2$ is chosen to be inversely proportional to the volume of the 3-balls $B_\pm^3$. From \eqref{6.18}-\eqref{6.24} we conclude that holomorphic $\Psi_+$ and antiholomorphic $\Psi_-$ solutions of the Klein-Gordon oscillator equation form weighted Bergman spaces defined as Hilbert spaces of square-integrable holomorphic functions on $B_\pm^3$ with measure $\mu_{N+2}$,
\begin{equation}\label{6.25}
\CH^\pm = L_h^2(B_\pm^3, \mu_{N+2}^{}):=\{\Psi_\pm^{}\in\CF^\pm\ \mbox{with}\ \eqref{6.18}-\eqref{6.24}\mid
\langle\Psi_\pm^{} , \Psi_\pm^{}\rangle<\infty\}\ .
\end{equation}
For more details on weighted Bergman spaces, see e.g. \cite{BH, ZZ} and references therein.

The basis in the weighted Bergman space \eqref{6.25} is given by functions
\begin{equation}\label{6.26}
f_\pm (n_1, n_2, n_3)=(y^1_\pm)^{n_1}(y^2_\pm)^{n_2}(y^3_\pm)^{n_3}\ ,\ n_a =0, 1, ...\ ,
\end{equation}
which yield eigenstates of the Klein-Gordon oscillator of the form
\begin{equation}\label{6.27}
\begin{split}
\Psi_\pm^{}(N,n_1,n_2,n_3)= 
\frac{e^{N/2}}{(\omega z_\pm^0)^{N+2}}(y^1_\pm)^{n_1}(y^2_\pm)^{n_2}(y^3_\pm)^{n_3}v_\pm^{}\\
=\frac{e^{N/2}}{\omega^{N+2}}\Bigl(\frac{1}{z_\pm^0}\Bigr)^{n_0}(z_\pm^1)^{n_1}(z_\pm^2)^{n_2}(z_\pm^3)^{n_3}v_\pm ,
\end{split}
\end{equation}
where $n_0:=n_1+n_2+n_3+N+2$. Note that the number 2 in \eqref{6.15} and \eqref{6.16} comes from the convolution of $\sfrac12\eta^{\mu\nb}$ with the commutator \eqref{5.15}, so the square of the energy operator $\hat E^2$ (associated with $z^0\zb^{\bar 0}$ in $H$) is 
\begin{equation}\label{6.28}
\hat E^2:=-\frac{2\hbar c^2}{w^2}\,\Bigl(z^0\dpar_{z^0} + \frac{1}{2}\Bigr)=\frac{2\hbar c^2}{w^2}\,\Bigl(z^a\dpar_{z^a}+\frac{3}{2} +\frac{c^2}{\hbar}N\Bigr )\ ,
\end{equation}
where we have restored $\hbar$ and $c$, and $w^{-2}=m\omega$.
The eigenvalues of the operator $\hat E$ are associated with the radius of the disk on the $(x^0, p_0)$-plane,
\begin{equation}\label{6.29}
\begin{split}
E(n_1, n_2, n_3)=\sqrt{\frac{2\hbar c^2}{w^2}(n_0{-}\sfrac12)} =mc^2\sqrt{1{+}\frac{2\hbar\omega}{mc^2}(n_1{+}n_2{+}n_3{+}\sfrac32)}\\
\cong  mc^2+\hbar\omega  (n_1+n_2+n_3+\sfrac32)\for c^2\to\infty\ ,\quad
\end{split}
\end{equation}
so they are positive for all states of particles $\Psi_+$ and antiparticles $\Psi_-$. 
Formula \eqref{6.29} shows that in the non-relativistic limit  $c^2\to\infty$ the difference $E(n_1, n_2, n_3)-mc^2$ is the energy of the $3D$ harmonic oscillator. Note that in this paper we did not care about normalization factors in eigenstates, integrals, etc., they are not important here.

\section{Conclusions}

\noindent 
A classical relativistic spin-0 particle of mass $m$ is defined by a point in phase space $T^*\R^{3,1}=\R^{3,1}\times \R^{3,1}$ with coordinates $x^\mu$ and momenta $p_\mu$. The dynamics are given by the choice of a Hamiltonian function $H$ (it is not energy) on $T^*\R^{3,1}$ whose constant value fixes a 7-dimensional hypersurface $X_7\subset T^*\R^{3,1}$ in phase space. This function $H$ also defines a Hamiltonian vector field $V_H$ generating a one-parameter group with elements $g=\exp(\tau V_H)$ acting on $X_7$. Here $\tau$ is a parameter on the orbit in $X_7$ along which the particle moves, and the space $X_6$ of all orbits  is obtained by quotienting $X_7$ by the action of this group. This manifold $X_6$ parametrizes initial data of the particle's motion and is called covariant phase space.

We have shown that the covariant phase space  of the classical relativistic oscillator is the homogeneous SU(3,1)-space
\begin{equation}\label{7.1}
Z_6=\sSU(3,1)/\rS(\sU(3)\times\sU(1))\ .
\end{equation}
On this Riemannian manifold there exist two almost complex structures $\CJ_\pm =\pm\CJ$. If we choose the initial momentum with $p^0>0$ in space $B_+^3=(Z_6, \CJ_+)$, then in space $B_-^3=(Z_6, \CJ_-)$ we will have $p^0<0$. Therefore, we identified $B_+^3$ as the space of initial data for particles and $B_-^3$ as the initial data space for antiparticles. In the limit when the frequency parameter $\omega$ tends to zero, these manifolds are deformed into cotangent bundles $T^*H_\pm^3$,
\begin{equation}\label{7.2}
B_\pm^3\quad\stackrel{{\omega\to 0}}{\longrightarrow}\quad T^*H_\pm^3\ ,
\end{equation}
defined by the equations
\begin{equation}\label{7.3}
T^*H_\pm^3:\quad \eta^{\mu\nu}p_\mu p_\nu + m^2=0\und 
p_\mu x^\mu =0\with \qv=\mbox{sgn}(p^0)=\pm 1\ \mbox{on}\ H_\pm^3\ .
\end{equation}
These spaces describe the initial data of the particle for $\qv =1$ and the antiparticle for $\qv =-1$. Interaction with an electromagnetic field is introduced by replacing $p_\mu$ with $P_\mu =p_\mu + e A_\mu$ in the function $H(x,p)$. As a result we obtain one-parametric family of 6-dimensional covariant phase spaces $Y_6=X_6(e)$ with deformation parameter $e$.

We described relativistic quantum mechanics of spinless particles as an Abelian gauge theory on the phase space $T^*\R^{3,1}$ basing ourselves on the ideas of the geometric quantization approach \cite{Sour}-\cite{Wood}. The main object of theory are covariant derivatives $\nabla_A$ containing a background gauge field $\Av =\theta_{\R^{6,2}}^{}J$, where $\theta_{\R^{6,2}}^{}$ is a potential of the symplectic 2-form $\omega_{\R^{6,2}}^{}=\dd\theta_{\R^{6,2}}^{}$ on $T^*\R^{3,1}$. Particles $\Psi_+$ and antiparticles $\Psi_-$ are defined through sections $\Psi =\Psi_++\Psi_-$ of the bundle $L_{\C^2}^{}=L_\C^+ \oplus L_\C^-$ over the phase space. The structure group of the complex conjugate bundles $L_\C^+$ and  $L_\C^-$ is the group U(1)$_{\sf v}$. The formulation in the standard language of gauge theories shows that particles $\Psi_+\in L_\C^+$ and antiparticles $\Psi_-\in L_\C^-$ are characterized by a new type of charge -- a quantum charge $\qv =\pm 1$ -- reflecting their interaction with the background Abelian gauge field $\Av$. Thus, from the point of view of differential geometry, the transition from classical to quantum mechanics corresponds to the replacement of partial derivatives on the phase space by covariant derivatives on the bundle $L_{\C^2}^{}$ over phase space and endowment of ``wavefunctions" $\Psi_\pm\in L_\C^\pm$ with a charge $\qv =\pm 1$ of the group U(1)$_{\sf v}$, that leads to the switching  on of the interaction of $\Psi_\pm$ with the gauge field $\Av$.

Having described the covariant phase space  \eqref{7.1} of the classical relativistic oscillator, we have moved on to first quantized theory. The most natural differential operator of gauge theory on phase space is the covariant Laplacian $\Delta_2$, and the eigenfunction problem of this operator leads to the Klein-Gordon oscillator equation. We have shown that solutions of this equation in the complex Segal-Bargmann representation are direct sum of holomorphic solutions $\Psi_+$ for particles and antiholomorphic solutions $\Psi_-$ for antiparticles, with the energy eigenstates forming weighted Bergman spaces
\begin{equation}\label{7.4}
\CH^\pm = L_h^2(B_\pm^3, \mu_{N+2}^{})
\end{equation}
of square-integrable holomorphic functions on covariant phase spaces $B_\pm^3$ of classical oscillators. These spaces $\CH^\pm$ are Hilbert spaces of unitary representation of the group SU(3,1) and its subgroup SO(3,1). Thus, the relativistic quantum harmonic oscillator is an exactly solvable unitary model that does not contain non-physical states. All information about classical oscillators with $\qv =\pm 1$ is contained in their covariant phase spaces $B_\pm^3$, and all information about quantum oscillators with $\qv =\pm 1$ is contained in the weighted Bergman spaces $\CH^\pm$ from \eqref{7.4}. Such correspondence holds for all single-particle systems. Further research is needed.

\bigskip

\noindent 
{\bf\large Acknowledgments}

\noindent
I am grateful to Tatiana Ivanova for stimulating discussions and remarks.


\end{document}